\documentclass[
prd,preprint,aps,endfloats,tightenlines,
showpacs,superscriptaddress,nofootinbib,
amssymb
]{revtex4}
\usepackage{graphicx}
\begin{document}
\setlength{\baselineskip}{16pt}
%
%
\title{
Lattice QCD Calculation of the $\rho$ Meson Decay Width
}
%
%
\author{ S.~Aoki }
\affiliation{
Graduate School of Pure and Applied Sciences,
University of Tsukuba,
Tsukuba, Ibaraki 305-8571, Japan
}
\affiliation{
Riken BNL Research Center,
Brookhaven National Laboratory,
Upton, New York 11973, USA
}
%
\author{ M.~Fukugita }
\affiliation{
Institute for Cosmic Ray Research,
University of Tokyo,
Kashiwa 277 8582, Japan
}
%
\author{ K-I.~Ishikawa }
\affiliation{
Department of Physics,
Hiroshima University,
Higashi-Hiroshima, Hiroshima 739-8526, Japan
}
%
\author{ N.~Ishizuka }
\affiliation{
Graduate School of Pure and Applied Sciences,
University of Tsukuba,
Tsukuba, Ibaraki 305-8571, Japan
}
\affiliation{
Center for Computational Sciences,
University of Tsukuba,
Tsukuba, Ibaraki 305-8577, Japan
}
%
\author{ K.~Kanaya }
\affiliation{
Graduate School of Pure and Applied Sciences,
University of Tsukuba,
Tsukuba, Ibaraki 305-8571, Japan
}
%
\author{ Y.~Kuramashi }
\affiliation{
Graduate School of Pure and Applied Sciences,
University of Tsukuba,
Tsukuba, Ibaraki 305-8571, Japan
}
\affiliation{
Center for Computational Sciences,
University of Tsukuba,
Tsukuba, Ibaraki 305-8577, Japan
}
%
\author{ Y.~Namekawa }
\altaffiliation[ Present address : ]{
Center for Computational Sciences,
University of Tsukuba,
Tsukuba, Ibaraki 305-8577, Japan
}
\affiliation{
Department of Physics,
Nagoya University,
Nagoya 464-8602, Japan
}
%
\author{ M.~Okawa }
\affiliation{
Department of Physics,
Hiroshima University,
Higashi-Hiroshima, Hiroshima 739-8526, Japan
}
%
\author{ K.~Sasaki }
\affiliation{
Center for Computational Sciences,
University of Tsukuba,
Tsukuba, Ibaraki 305-8577, Japan
}
%
\author{ A.~Ukawa }
\affiliation{
Graduate School of Pure and Applied Sciences,
University of Tsukuba,
Tsukuba, Ibaraki 305-8571, Japan
}
\affiliation{
Center for Computational Sciences,
University of Tsukuba,
Tsukuba, Ibaraki 305-8577, Japan
}
%
\author{  T.~Yoshi\'{e} }
\affiliation{
Graduate School of Pure and Applied Sciences,
University of Tsukuba,
Tsukuba, Ibaraki 305-8571, Japan
}
\affiliation{
Center for Computational Sciences,
University of Tsukuba,
Tsukuba, Ibaraki 305-8577, Japan
}
%
\collaboration{ CP-PACS Collaboration }
%
%
\date{ \today }
%
%
\begin{abstract}
We present a lattice QCD calculation of the $\rho$ meson decay width
via the $P$-wave scattering phase shift for the $I=1$ two-pion system.
Our calculation uses full QCD gauge configurations
for $N_f=2$ flavors generated
using a renormalization group improved gauge action
and an improved Wilson fermion action on a $12^3\times24$ lattice
at $m_\pi/m_\rho=0.41$
and the lattice spacing $1/a=0.92\ {\rm GeV}$.
The phase shift calculated with the use of the finite size formula
for the two-pion system in the moving frame
shows a behavior consistent with the existence of a resonance
at a mass close to the vector meson mass
obtained in spectroscopy.
The decay width estimated from the phase shift
is consistent with the experiment,
when the quark mass is scaled to the realistic value.
\end{abstract}
\maketitle
%
%
\section{ Introduction }
\label{Sec:Introduction}
A study of the $\rho$ meson decay
is a significant step for understanding
the dynamical aspect of hadron reactions with lattice QCD.
We notice three studies carried out to date
toward this direction.
The earlier two~\cite{rhpipi_GMTW,rhpipi_LD}
employed the quenched approximation
ignoring the decay into two ghost pions
which appear in the quenched theory.
The third one used the full QCD~\cite{rhpipi_MM}
and estimated the decay width
from the $\rho\to\pi\pi$ transition amplitude
$\langle \rho | \pi\pi \rangle$
extracted from the time behavior of the correlation functions
$\langle 0 | \pi(t)\pi(t) \rho(0)      | 0 \rangle$ and
$\langle 0 | \pi(t)\pi(t) \pi(0)\pi(0) | 0 \rangle$,
assuming that the hadron interaction is small.
These studies, however,
were all carried out at unphysical kinematics
$m_\pi/m_\rho > 1/2$.

In the present work we attempt to carry out
a more realistic calculation.
We estimate the $\rho$ meson decay width
by calculating the $P$-wave scattering phase shift
for the $I=1$ two-pion system.
The calculations are carried out with
$N_f=2$ full QCD configurations previously generated
for a study of the light hadron spectrum
with a renormalization group improved gauge action
and a clover fermion action at $\beta=1.8$, $\kappa=0.14705$
on a $12^3\times 24$ lattice~\cite{conf_NMC}.
The lattice parameters were determined from the spectrum analysis
which gave $m_\pi/m_\rho=0.41$, the lattice extent $L=2.53\ {\rm fm}$
and the lattice space inverse $1/a=0.92\ {\rm GeV}$.
The finite size formula
presented by Rummukainen and Gottlieb~\cite{fm_RG}
is employed to estimate the phase shift.
This calculation is made at two energies
that allow to study the existence of the resonance.

This paper is organized as follows.
In Sec.~\ref{Sec:Methods}
we give the method of the calculations
and the simulation parameters.
We present our results in Sec.~\ref{Sec:Results}.
Our conclusions are given in Sec.~\ref{Sec:Conclusions}.
Preliminary reports of the present work
were presented in~\cite{rhpipi_CP-PACS}.
The calculation was carried out
on VPP5000/80 at the Academic Computing
and Communications Center of University of Tsukuba.
%
%
\section{ Methods }
\label{Sec:Methods}
%
\subsection{ Rummukainen-Gottlieb formula }
\label{SubSec:Rummukainen-Gottlieb formula}
Let us consider the $\rho$ meson decay
into the two pions in the $P$-wave.
When the $\rho$ meson is at rest,
the energy of the two pions,
neglecting the final state interaction, is
\begin{equation}
   E = 2 \sqrt{ m_{\pi }^2 + p^2 }
\ ,
\label{eq:E_CM}
\end{equation}
where
the momentum takes
${\bf p}=(2\pi/L) {\bf n}$ (${\bf n}\in \mathbb{Z}^3 \not= 0$)
on the lattice.
In a typical full QCD simulation on $L\sim 2-3\ {\rm fm}$,
this energy is significantly larger than the resonance mass $m_\rho$.
On our full QCD configurations, for example,
the lowest energy estimated from $m_\pi$ and $m_\rho$
calculated in the previous study~\cite{conf_NMC}
is $E=1.47\times m_\rho$, appreciably away from the
resonance and it is not suitable to study the $\rho$ meson decay.

In order to realize the kinematics such that the energy
of the two pions is close to $m_\rho$,
we consider a system having a non-zero total momentum,
{\it i.e.,} the moving frame~\cite{fm_RG},
with the total momentum
${\bf p}=p {\bf e}_3 = (2\pi / L) {\bf e}_3$
in a box satisfying the $L^3$ periodic boundary condition.
We set the system to the ${\bf A}_2^-$ representation
of the rotation group on the lattice
(the tetragonal rotation group ${\rm D}_{4h}$),
which represents the $J=1$ spin state
ignoring effects from higher spin states with $J \geq 3$.
We consider the iso-spin representation of $(I,I_z)=(1,0)$,
the neutral $\rho$ meson.

For non-interacting hadrons
the dominant low energy states in the moving frame
are the two free pions with the momenta ${\bf p}$ and ${\bf 0}$,
and the $\rho$ meson with the momentum ${\bf p}$
and a polarization vector parallel to the momentum, $\rho_3({\bf p})$.
The energies of these states are
\begin{equation}
\begin{array}[b]{lll}
   W_1^0 = \sqrt{ m_{\pi }^2 + p^2 } + m_\pi
            && \quad \mbox{ for the two free pions }
\ , \\
   W_2^0 = \sqrt{ m_{\rho}^2 + p^2 }
            && \quad \mbox{ for the $\rho$ meson   }
\ .
\label{eq:W0_moving}
\end{array}
\end{equation}
Other states having higher energies are neglected.
On our full QCD configurations
the invariant mass of the two free pions
takes $\sqrt{s} = 0.97\times m_\rho$,
which is closer to $m_\rho$
than that given by (\ref{eq:E_CM})
for the system having the zero total momentum.

The hadron interaction shifts the energy from $W_n^0$ to $W_n$ ($n=1,2$),
and the energies $W_n$ are related to the two-pion scattering phase shift
$\delta$ in the infinite volume
through the Rummukainen-Gottlieb formula~\cite{fm_RG},
which is an extension of the L\"uscher formula~\cite{fm_LU}
to the moving frame.
The formula for the ${\bf A}_2^-$ representation
and the total momentum ${\bf p}=p{\bf e}_3$ reads
\begin{equation}
  \frac{1}{\tan\delta} = Z( 1 ; k L / (2\pi) )
\ ,
\label{eq:RG_F}
\end{equation}
where $k$ is the momentum
defined from the invariant mass $\sqrt{s}$ as
$\sqrt{s} = \sqrt{ W^2 - p^2 } = 2 \sqrt{ k^2 + m_\pi^2 }$.
The function $Z$ is an analytic continuation of
\begin{equation}
  Z( x ; q )
  = \frac{ 1 }{ 2\pi^2 q \gamma }
    \sum_{ {\bf r} \in \Gamma }
       \frac{ 1 + ( 3 r_3^2 - r^2 )/q^2 }
            { ( r^2 - q^2 )^{x} }
\ ,
\label{eq:RG_F_two}
\end{equation}
which is defined for ${\rm Re}(x) > 5/2$,
where $\gamma=W/\sqrt{s}$ is the Lorentz boost factor
and the summation for ${\bf r}$ runs over the set
\begin{equation}
  \Gamma =
    \Biggl\{
     \ {\bf r} \ |
     \ r_1 = n_1 \ ,
     \ r_2 = n_2 \ ,
     \ r_3 = \Biggl( n_3 + \frac{p}{2} \frac{L}{2\pi} \Biggr)/\gamma \ ,
     \ {\bf n}\in \mathbb{Z}^3
     \ \Biggr\}
\ .
\end{equation}
$Z(1;q)$ can be evaluated
by the method described in Ref.~\cite{phsh_YAMA}.
%
\subsection{ Extraction of energies }
\label{SubSec:Extraction of energies}
In order to calculate the two energies $W_n$ ($n=1,2$)
we construct a matrix of the time correlation function,
\begin{equation}
  G(t) =
  \left(
  \begin{array}{ll}
     \     \langle 0 | \ (\pi\pi)^\dagger(t) \ (\pi\pi) (t_S) \ | 0 \rangle
&    \quad \langle 0 | \ (\pi\pi)^\dagger(t) \   \rho_3 (t_S) \ | 0 \rangle  \\
     \     \langle 0 | \   \rho_3^\dagger(t) \ (\pi\pi) (t_S) \ | 0 \rangle
&    \quad \langle 0 | \   \rho_3^\dagger(t) \   \rho_3 (t_S) \ | 0 \rangle
  \end{array}
  \right)
\ ,
\label{eq:G}
\end{equation}
where $\rho_3(t)$ is an interpolating operator
for the neutral $\rho$ meson with the momentum
${\bf p}=(2\pi/L) {\bf e}_3$
and the polarization vector parallel to ${\bf p}$,
and $(\pi\pi)(t)$ is an interpolating operator
for the two free pions,
\begin{equation}
  (\pi\pi)(t) = \frac{1}{\sqrt{2}}
     \Bigl(   \pi^{-}({\bf p},t) \pi^{+}({\bf 0},t)
            - \pi^{+}({\bf p},t) \pi^{-}({\bf 0},t)   \Bigl)
\ .
\label{eq:pp_op}
\end{equation}
These operators belong to the ${\bf A}_2^-$
and the $(I,I_z)=(1,0)$.

To extract $W_n$ ($n=1,2$)
we construct a matrix,
\begin{equation}
  M(t,t_R) = G(t) G^{-1}(t_R)
\ ,
\label{eq:M_def}
\end{equation}
with some reference time $t_R$~\cite{method_diag}.
The two eigenvalues $\lambda_n (t,t_R)$ ($n=1,2$)
of the matrix $M(t,t_R)$ behave as
\begin{equation}
   \lambda_n (t,t_R) = {\rm e}^{ - W_n \cdot ( t - t_R ) }
\ ,
\end{equation}
for a large $t$,
if the two lowest states dominate the correlation function.
The two energies $W_n$ can be extracted
by a single exponential fit to $\lambda_n(t,t_R)$.

In order to construct the meson state with a non-zero momentum
we introduce $U(1)$ noise $\xi_j({\bf x})$
in three-dimensional space,
which satisfies
\begin{equation}
  \sum_{j=1}^{N_R}
       \xi_j^* ({\bf x}) \xi_j ({\bf y})
          = \delta^3 ( {\bf x} - {\bf y} )
       \qquad \mbox{ for \ $N_R \to \infty$ }
\ ,
\label{eq:xi_prop}
\end{equation}
where $N_R$ is the number of the noise representation taken
to be $10$ in the present work.
We calculate the quark propagator,
\begin{equation}
  Q_{AB}( {\bf x}, t | {\bf q}, t_S, \xi_j )
    = \sum_{\bf y} ( D^{-1} )_{AB}({\bf x}, t ; {\bf y}, t_S )
      \cdot \Bigl[ {\rm e}^{ i{\bf q}\cdot{\bf y}} \xi_j({\bf y}) \Bigr]
\ ,
\label{eq:QP_Q}
\end{equation}
where $A$ and $B$ refer to color and spin indices.
The square bracket in (\ref{eq:QP_Q})
is taken as the source term in solving the propagator.
The two point function of the meson with the spin content $\Gamma$
and the momentum ${\bf p}$ can be constructed from $Q$ by
\begin{equation}
  \sum_{j=1}^{N_R} \
  \sum_{\bf x} {\rm e}^{ - i {\bf p}\cdot {\bf x} }
  \ \Bigl\langle
      \gamma_5
         Q^\dagger ( {\bf x}, t | {\bf 0}, t_S, \xi_j ) \gamma_5 \Gamma^\dagger
         Q         ( {\bf x}, t | {\bf p}, t_S, \xi_j )          \Gamma
    \Bigr\rangle
\ ,
\end{equation}
where the bracket means the trace
with respect to the color and the spin indices;
hereafter we take this convention for the brackets.

\begin{figure}[h]
\includegraphics[width=14.5cm]{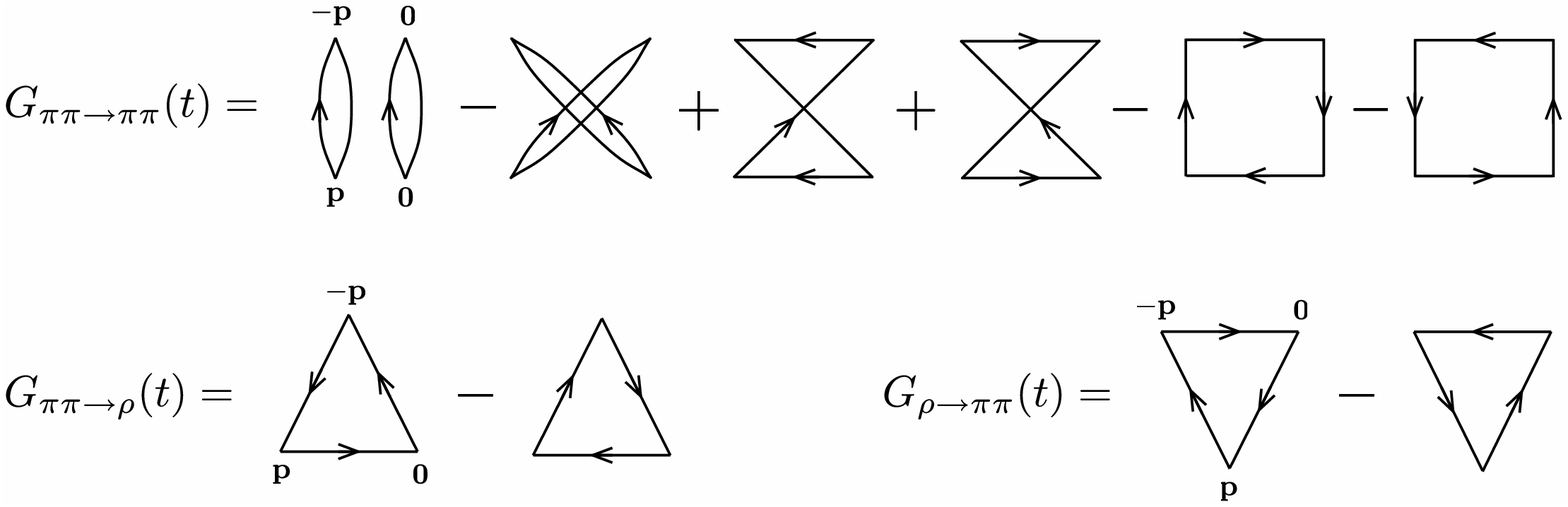}
\caption{
Quark contractions of
$\pi\pi\to\pi\pi$, $\pi\pi\to\rho$ and
$\rho\to\pi\pi$ components of
the time correlation function $G(t)$.
Vertices refer to the pion or the $\rho$ meson
with a momentum specified in the diagram.
The time runs 
upward in the diagrams.
}
\label{fig:quark_c}
\newpage
\end{figure}

The contractions of the quark field for the components of $G(t)$
are shown in Fig.~\ref{fig:quark_c}.
The vertices refer to the pion or the $\rho$ meson
with the momentum specified in the diagrams.
The time runs upward in the diagrams.
The function $G_{\pi\pi\to\pi\pi}(t)$ for the first diagram
in Fig.~\ref{fig:quark_c} is calculated
by introducing in addition another $U(1)$ noise $\eta_j({\bf x})$
having the property identical to that of $\xi_j({\bf x})$
as in (\ref{eq:xi_prop}),
\begin{equation}
  G_{\pi\pi\to\pi\pi}^{\rm [1st]} =
  \sum_{j=1}^{N_R}
  \sum_{{\bf x}, {\bf y}} {\rm e}^{ - i {\bf p}\cdot {\bf x} }
  \ \Bigl\langle
       Q^\dagger ( {\bf x}, t | {\bf 0}, t_S, \xi_j  )
       Q         ( {\bf x}, t | {\bf p}, t_S, \xi_j  ) \Bigr\rangle
    \Bigl\langle
       Q^\dagger ( {\bf y}, t | {\bf 0}, t_S, \eta_j )
       Q         ( {\bf y}, t | {\bf 0}, t_S, \eta_j ) \Bigr\rangle
\ .
\label{eq:D1_quark_c}
\end{equation}
The function $G_{\pi\pi\to\pi\pi}(t)$ for the second diagram
is obtained by exchanging the momenta of the sink
in (\ref{eq:D1_quark_c}).

To obtain the $G(t)$ for the other diagrams
we calculate the quark propagator of a different type
by the source method,
\begin{equation}
  W_{AB} ( {\bf x}, t | {\bf k}, t_1 | {\bf q}, t_S, \xi_j )
  = \sum_{\bf z}
    \sum_{C}
       ( D^{-1} )_{AC}({\bf x}, t ; {\bf z}, t_1 )
         \cdot
         \Bigl[ {\rm e}^{ i{\bf k}\cdot{\bf z}}
                      \gamma_5 \ Q( {\bf z}, t_1 | {\bf q}, t_S, \xi_j )
         \Bigr]_{CB}
\ ,
\label{eq:QP_W}
\end{equation}
where $A$, $B$ and $C$ refer to color and spin indices,
and the term in the square bracket is taken
as the source term in solving the propagator.
Using $W$
we calculate the functions $G_{\pi\pi\to\pi\pi}(t)$
for the third to sixth diagrams in Fig.~\ref{fig:quark_c} by
\begin{eqnarray}
  G_{\pi\pi\to\pi\pi}^{\rm [3rd]} &=&
  \sum_{j=1}^{N_R}
  \sum_{\bf x}  {\rm e}^{ - i {\bf p}\cdot {\bf x} }
  \ \Bigl\langle
       W^\dagger ( {\bf x}, t | {\bf 0}, t_S | - {\bf p}, t_S, \xi_j )
       W         ( {\bf x}, t | {\bf 0}, t   |   {\bf 0}, t_S, \xi_j )
    \Bigr\rangle
\ ,
\cr
  G_{\pi\pi\to\pi\pi}^{\rm [4th]} &=&
  \sum_{j=1}^{N_R}
  \sum_{\bf x}  {\rm e}^{ - i {\bf p}\cdot {\bf x} }
  \ \Bigl\langle
       W         ( {\bf x}, t | {\bf 0}, t_S |   {\bf p}, t_S, \xi_j )
       W^\dagger ( {\bf x}, t | {\bf 0}, t   |   {\bf 0}, t_S, \xi_j )
    \Bigr\rangle
\ ,
\cr
  G_{\pi\pi\to\pi\pi}^{\rm [5th]} &=&
  \sum_{j=1}^{N_R}
  \sum_{\bf x}  {\rm e}^{ - i {\bf p}\cdot {\bf x} }
  \ \Bigl\langle
       W         ( {\bf x}, t | {\bf p}, t_S | {\bf 0}, t_S, \xi_j )
       W^\dagger ( {\bf x}, t | {\bf 0}, t   | {\bf 0}, t_S, \xi_j )
    \Bigr\rangle
\ ,
\cr
  G_{\pi\pi\to\pi\pi}^{\rm [6th]} &=&
  \sum_{j=1}^{N_R}
  \sum_{\bf x}  {\rm e}^{ - i {\bf p}\cdot {\bf x} }
  \ \Bigl\langle
       W^\dagger ( {\bf x}, t | - {\bf p}, t_S | {\bf 0}, t_S, \xi_j )
       W         ( {\bf x}, t |   {\bf 0}, t   | {\bf 0}, t_S, \xi_j )
    \Bigr\rangle
\ .
\label{eq:QC_B}
\end{eqnarray}
The functions $G_{\pi\pi\to\rho}(t)$
for the two diagrams in Fig.~\ref{fig:quark_c}
can be calculated by
\begin{eqnarray}
  G_{\pi\pi\to\rho}^{\rm [1st]} &=&
  \sum_{j=1}^{N_R}
  \sum_{\bf x}  {\rm e}^{ - i {\bf p}\cdot {\bf x} }
  \ \Bigl\langle
       Q         ( {\bf x}, t                  | {\bf 0}, t_S, \xi_j )
       W^\dagger ( {\bf x}, t | - {\bf p}, t_S | {\bf 0}, t_S, \xi_j )
       \gamma_5 \gamma_3
    \Bigr\rangle
\ ,
\cr
  G_{\pi\pi\to\rho}^{\rm [2nd]} &=&
  \sum_{j=1}^{N_R}
  \sum_{\bf x}  {\rm e}^{ - i {\bf p}\cdot {\bf x} }
  \ \Bigl\langle
       W         ( {\bf x}, t | {\bf p}, t_S | {\bf 0}, t_S, \xi_j )
       Q^\dagger ( {\bf x}, t                | {\bf 0}, t_S, \xi_j )
       \gamma_5 \gamma_3
    \Bigr\rangle
\ ,
\label{eq:QC_pprh}
\end{eqnarray}
and similarly for $G_{\rho\to\pi\pi}(t)$,
\begin{eqnarray}
  G_{\rho\to\pi\pi}^{\rm [1st]} &=&
  - \sum_{j=1}^{N_R}
  \sum_{\bf x}  {\rm e}^{ - i {\bf p}\cdot {\bf x} }
  \ \Bigl\langle
       W^\dagger ( {\bf x}, t | {\bf 0}, t | {\bf 0}, t_S, \xi_j )
       Q         ( {\bf x}, t              | {\bf p}, t_S, \xi_j )
       \gamma_5 \gamma_3
    \Bigr\rangle
\ ,
\cr
  G_{\rho\to\pi\pi}^{\rm [2nd]} &=&
  - \sum_{j=1}^{N_R}
  \sum_{\bf x}  {\rm e}^{ - i {\bf p}\cdot {\bf x} }
  \ \Bigl\langle
       Q^\dagger ( {\bf x}, t              | -{\bf p}, t_S, \xi_j )
       W         ( {\bf x}, t | {\bf 0}, t |  {\bf 0}, t_S, \xi_j )
       \gamma_5 \gamma_3
    \Bigr\rangle
\ .
\label{eq:QC_rhpp}
\end{eqnarray}

The quark propagators are solved
with the Dirichlet boundary condition
imposed in the time direction
and the source operator is set at $t_S=4$
which is sufficiently large
to avoid effects from the temporal boundary.
We calculate the $Q$-type propagators (\ref{eq:QP_Q})
for four sets of ${\bf q}$ and the $U(1)$ noise:
\begin{equation}
  ( {\bf q}, {\rm noise})=\{
\ ( {\bf 0}, \xi  ) ,
\ ( {\bf 0}, \eta ) ,
\ ( {\bf p}, \xi  ) ,
\ (-{\bf p}, \xi  )
\ \}
\ .
\end{equation}
The $W$-type propagators (\ref{eq:QP_W})
are calculated for 22 sets of ${\bf k}$, $t_1$ and ${\bf q}$:
\begin{equation}
  (  {\bf k}, t_1      |  {\bf q} )=\{
\ (  {\bf p}, t_S      |  {\bf 0} ) ,
\ ( -{\bf p}, t_S      |  {\bf 0} ) ,
\ (  {\bf 0}, t_S      |  {\bf p} ) ,
\ (  {\bf 0}, t_S      | -{\bf p} ) ,
\ (  {\bf 0}, t_1=4-21 |  {\bf 0} )
\ \}
\ ,
\end{equation}
using the same $U(1)$ noise representation $\xi$ in common.
All time correlation functions
can be calculated with combinations of these propagators.
We carry out additional measurements to reduce statistical errors
using the source operator located at $t_S+T/2$
with the Dirichlet boundary condition at $T/2$,
and average over the two measurements.
Thus we calculate $( 4 + 22 ) \times 10 \times 2 = 520$
quark propagators for each configuration.
%
\subsection{ Simulation parameters }
\label{SubSec:Simulation Parameters}
Calculations
employ $N_f=2$ full QCD configurations
previously generated for a study of the light hadron spectrum
using a renormalization group improved gauge action
and a clover fermion action
at $\beta=1.8$, $\kappa=0.14705$
with the mean-field improvement taking $C_{SW}=1.60$
on a $12^3\times 24$ lattice~\cite{conf_NMC}.
The periodic boundary conditions are imposed
for both spatial and temporal directions
in configuration generations
and the Dirichlet boundary condition
for the temporal direction
in calculations of quark propagators.
The lattice parameters determined from the spectrum analysis are
$m_\pi/m_\rho=0.41$, $L=2.53\ {\rm fm}$ and $1/a=0.92\ {\rm GeV}$.
The total number of configurations
analyzed every $5$ trajectories is $800$.
We estimate the statistical errors by the jackknife method
with bins of $100$ trajectories.
%
%
\section{ Results }
\label{Sec:Results}
%
\subsection{ Time correlation function }
\label{SubSec:Time Correlation Function}
\begin{figure}[h]
\includegraphics[width=11.0cm]{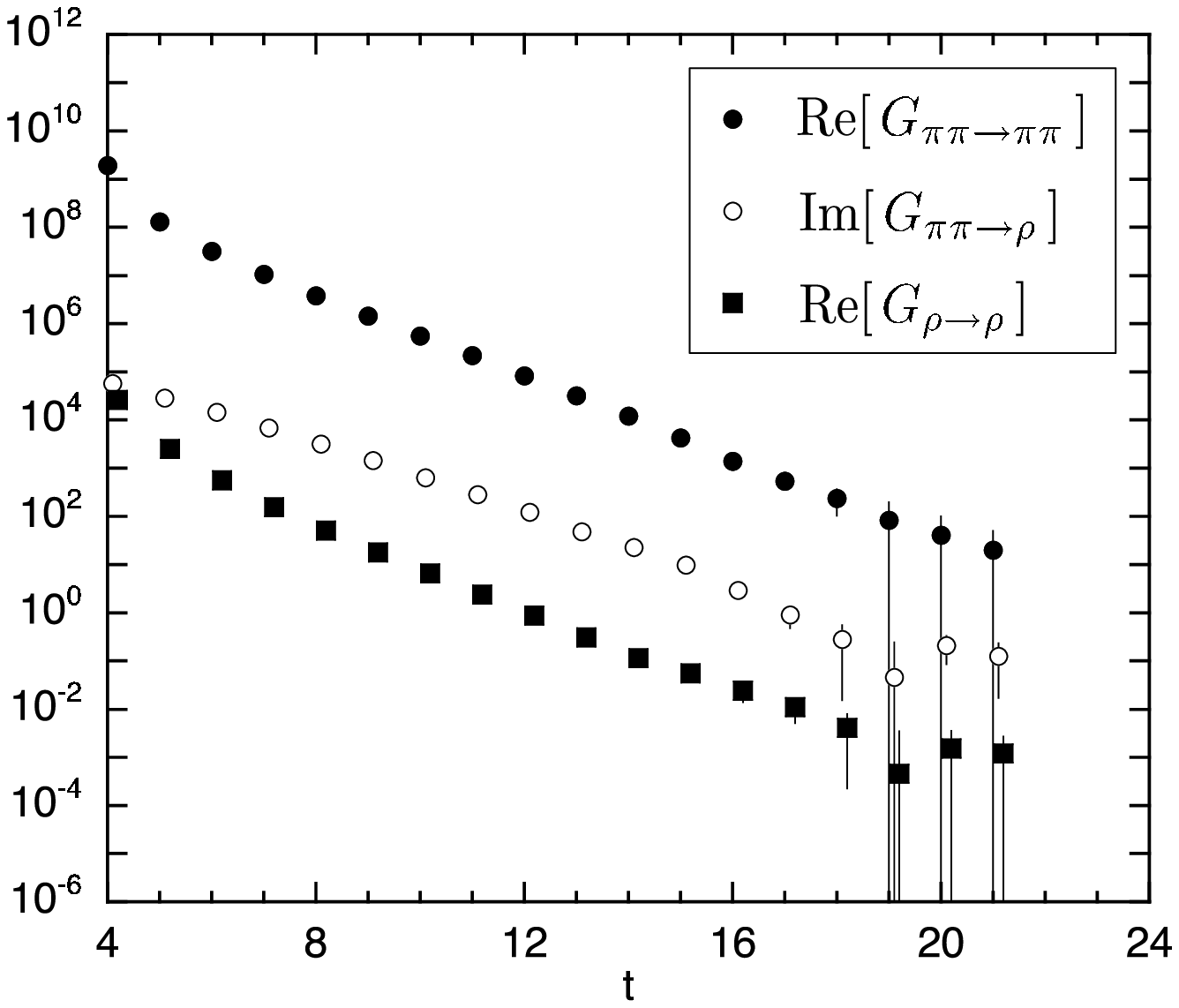} \\
\includegraphics[width=11.0cm]{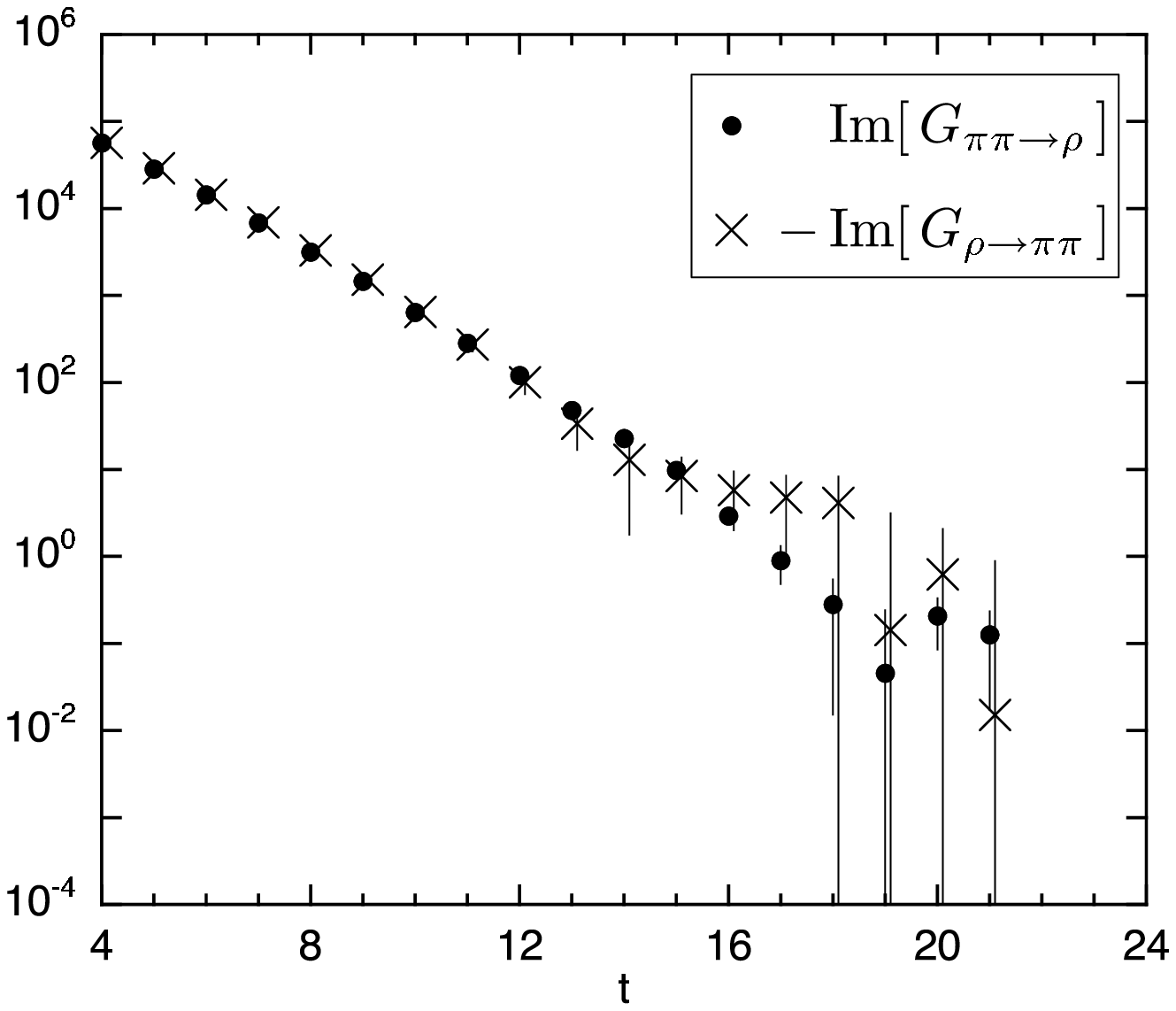}
\caption{
Real part of the diagonal components
($\pi\pi\to\pi\pi$ and $\rho\to\rho$)
and the imaginary part of the off-diagonal components
($\pi\pi\to\rho$ and $\rho\to\pi\pi$)
of the time correlation function $G(t)$.
}
\label{fig:G_t}
\newpage
\end{figure}

In Fig.~\ref{fig:G_t}
we show the real part of the diagonal components
($\pi\pi\to\pi\pi$ and $\rho\to\rho$)
and the imaginary part of the off-diagonal components
($\pi\pi\to\rho$ and $\rho\to\pi\pi$)
of the time correlation function $G(t)$ in (\ref{eq:G}).
$G(t)$ is a Hermitian matrix,
since the sink and source operators are identical
for a sufficiently large $N_R$
or equivalently for a large number of configurations.
The off-diagonal components are pure imaginary
by $P$ and $CP$ symmetry.
We find that these hold true within statistics.
The $\rho\to\pi\pi$ component agrees with $\pi\pi\to\rho$
as seen in Fig.~\ref{fig:G_t} within the error,
but the statistical errors of the former is large for a large $t$.
Hence, in the following analysis
we substitute $\rho\to\pi\pi$ by $\pi\pi\to\rho$
to reduce errors.

\begin{figure}[h]
\includegraphics[width=11.0cm]{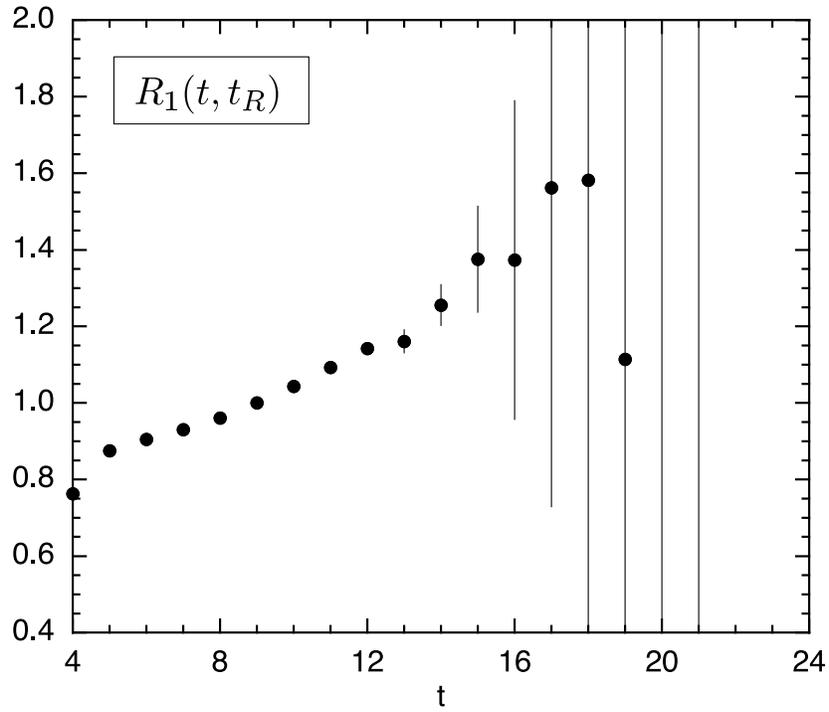} \\
\includegraphics[width=11.0cm]{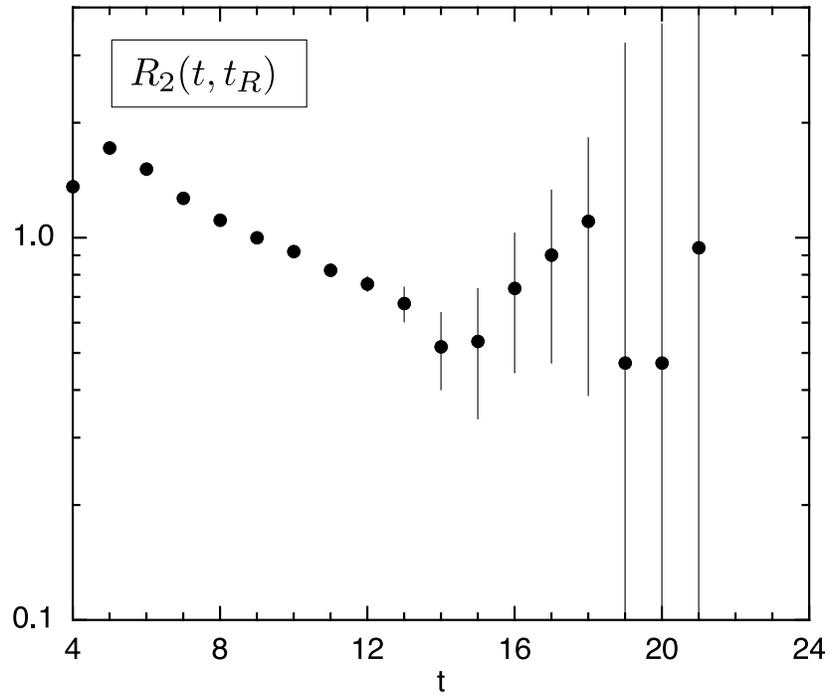}
\caption{
Normalized eigenvalues $R_1(t,t_R)$ and $R_2(t,t_R)$.
}
\label{fig:Lambda_t}
\newpage
\end{figure}

We calculate
the two eigenvalues $\lambda_n(t,t_R)$ ($n=1,2$)
for the matrix $M(t,t_R)$ in (\ref{eq:M_def})
with the reference time $t_R=9$.
In Fig.~\ref{fig:Lambda_t}
we plot the normalized eigenvalues $R_n(t,t_R)$ ($n=1,2)$
defined by
\begin{equation}
R_n(t,t_R)
= \lambda_n(t,t_R)
  \frac{ G_\pi (t_R;{\bf p}) \ G_\pi (t_R;{\bf 0}) }
       { G_\pi (t  ;{\bf p}) \ G_\pi (t  ;{\bf 0}) }
\ ,
\end{equation}
where $G_\pi (t;{\bf p})$ is the time correlation function
for the pion with the momentum ${\bf p}$,
\begin{equation}
G_\pi (t;{\bf p})
=\langle 0| \pi^\dagger ({\bf p},t) \pi({\bf p},t_S) |0\rangle
\ .
\label{eq:Gpi}
\end{equation}
The slope of the curves in Fig.~\ref{fig:Lambda_t}
represents the energy difference
with respect to the energy of the two free pions,
{\it i.e.}, $\Delta W_n = W_n - W_1^0$.
We observe that the energy difference
for $R_1(t,t_R)$ is negative
and that for $R_2(t,t_R)$ positive,
meaning that the phase shift
is positive and negative, respectively,
consistent with the presence of a resonance in between.

We extract the energy difference $\Delta W_n$ for both states
by a single exponential fit to
$R_n(t,t_R)$ for the time range $t=10-16$.
The energy of the two free pions $W_1^0$
is calculated from the mass $m_\pi$ and the energy $E$
obtained by a single exponential fit to
$G_\pi (t;{\bf 0})$ and $G_\pi (t;{\bf p})$ in (\ref{eq:Gpi}),
as $W_1^0=m_\pi + E$.
The energy $W_n$ is reconstructed
by $W_n=\Delta W_n + W_1^0$.
The results are tabulated
in the upper part of Table~\ref{table:TanDel}.
%
\subsection{ Effect of finite lattice spacing }
\label{SubSec: Effect of finite lattice spacing }
To see the size of errors arising from the discretization
of the energy and the momentum on the lattice,
we study the accuracy of the dispersion relation for the single particle.
In the upper part of Table~\ref{table:Disp}
we show the mass $m$ and the energy $E$
of the pion and the $\rho$ meson
with the momentum ${\bf p}=(2\pi/L){\bf e}_3$,
calculated from the time correlation functions.
The $\rho$ meson with zero momentum $\rho_j ({\bf 0})$ ($j=1,2,3$)
and that having the momentum with the perpendicular polarization
$\rho_j ({\bf p})$ ($j=1,2$) cannot decay energetically,
so that the mass and the energy
are obtained from the time correlation functions in a usual way.

In the continuum we have a relation
\begin{equation}
  E = \sqrt{ p^2 + m^2 }
\ ,
\label{eq:Disp_One_Cont}
\end{equation}
for the single particle.
We expect that the relation is modified on the lattice to
\begin{equation}
  {\rm cosh}( E ) = 2 \cdot \sin^2 ( p/2 ) + {\rm cosh}( m )
\ .
\label{eq:Disp_One_Lat}
\end{equation}
For the measured $m$ and $E$
we calculate the momenta $p_{\rm eff}$
from the two dispersion relations
(\ref{eq:Disp_One_Cont}) and (\ref{eq:Disp_One_Lat}).
We expect that these $p_{\rm eff}$
agree with the pre-set momentum $p=2\pi/L$
up to the discretization error.
The lower part of Table~\ref{table:Disp}
shows $p_{\rm eff}$ and the ratio $p_{\rm eff}^2 / p^2$
whose departure from unity is a measure
for the violation of the dispersion relation
due to the discretization error.
We find that the relation on the lattice (\ref{eq:Disp_One_Lat})
satisfies well,
whereas that in the continuum (\ref{eq:Disp_One_Cont})
does not hold so well for the pion.
For the $\rho$ meson
the departure from unity is not clearly detected
due to a large statistical error.

We should also be concerned
with the discretization error for the two-pion system.
We may think of two sources of the errors
in the Rummukainen-Gottlieb formula (\ref{eq:RG_F}).
One of them arises from the Lorentz transformation
from the moving frame to the center of mass frame
using Lorentz symmetry in the continuum.
In the transformation we use the relations,
\begin{eqnarray}
  \sqrt{s} & = & \sqrt{ W^2 - p^2 }  \ , \cr
       k^2 & = & s/4 - m_\pi^2       \ ,
\label{eq:Disp_Two_Cont}
\end{eqnarray}
for the invariant mass $\sqrt{s}$,
the energy in the moving frame $W$ and the momentum $k$.
These relations suffer from the discretization error,
and hence the definitions of $\sqrt{s}$ and $k$
contain similar errors.

After the Lorentz transformation
we obtain the Helmholtz equation
for the two-pion wave function $\phi({\bf x})$
in the center of mass frame,
\begin{equation}
   \Bigl( \nabla^2 + k^2 \Bigr) \phi({\bf x}) = 0
   \qquad \mbox{ for $|\vec{x}| > R $}
\ ,
\label{eq:H_eq}
\end{equation}
where $R$ is the two-pion interaction range.
In Ref.~\cite{fm_RG}
Rummukainen and Gottlieb derived the formula
by solving (\ref{eq:H_eq})
ignoring the effect of the finite lattice spacing.
Thus the discretization error also appears here.

The violation of the dispersion relation
in the continuum (\ref{eq:Disp_One_Cont})
has also been shown in Ref.~\cite{fm_RG},
where the phase shift for the $S$-wave state
was calculated for a statistical model.
Motivated from the validity of the dispersion relation
on the lattice (\ref{eq:Disp_One_Lat}),
Rummukainen and Gottlieb obtained
the invariant mass $\sqrt{s}$ and the momentum $k$
from the energy in the moving frame $W$ as
\begin{eqnarray}
  {\rm cosh}( \sqrt{s} ) &=& {\rm cosh}(W) - 2\cdot \sin^2(p/2)          \ , \cr
  2 \cdot \sin^2(k/2)    &=& {\rm cosh }(\sqrt{s}/2) - {\rm cosh}(m_\pi) \ ,
\label{eq:Disp_Two_Lat}
\end{eqnarray}
and the phase shift was obtained
by substituting $k$ into their formula.
The discretization error arising in solving (\ref{eq:H_eq}),
however, was not worked out.

In the present work
we calculate the invariant mass $\sqrt{s}$
and the momentum $k$
from the energy-momentum relations
both in the continuum (\ref{eq:Disp_Two_Cont})
and on the lattice (\ref{eq:Disp_Two_Lat}),
and estimate the phase shift
by putting $k$ into (\ref{eq:RG_F}).
We take the difference
arising from the two choices as the discretization error.
It is expected that this error vanishes in the continuum limit.
%
\subsection{ Scattering phase shift and decay width }
\label{SubSec:Scattering Phase Shift and Decay Width }
\begin{figure}[h]
\includegraphics[width=11.0cm]{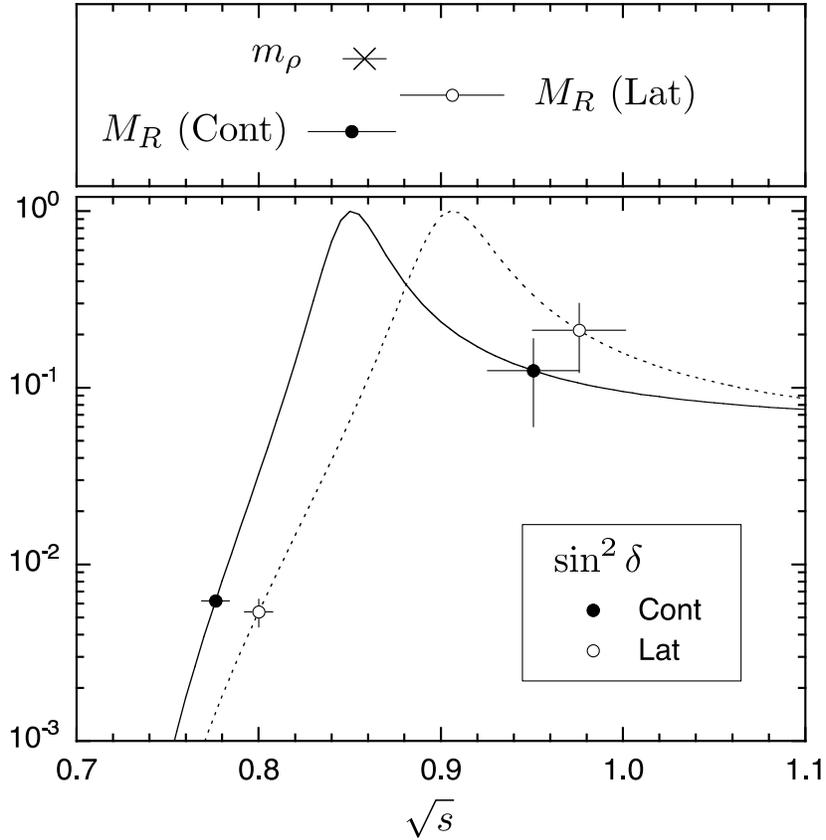}
\caption{
Scattering phase shift $\sin^2\delta$ ( lower panel ),
positions of $m_\rho$ and resonance mass $M_R$ ( upper panel ).
{\bf Cont} refer to the results
obtained with the energy-momentum relations
in the continuum (\ref{eq:Disp_Two_Cont})
and {\bf Lat} to those with the relations
on the lattice (\ref{eq:Disp_Two_Lat}).
The two lines are given by (\ref{eq:tanD_g})
with parameters $g_{\rho\pi\pi}$ and $M_R$
given in (\ref{eq:FinalR_Cont}) and (\ref{eq:FinalR_Lat}).
The abscissa is in lattice units.
}
\label{fig:sin2Del}
\newpage
\end{figure}

The invariant mass $\sqrt{s}$,
the momentum $k$ and the phase shift $\delta$
obtained by the procedure in the previous subsection
are presented in the lower part of Table~\ref{table:TanDel}.
Appreciable differences that depend on the energy-momentum relations used
are visible in $\sqrt{s}$ and $k$,
but the difference for $\delta$ is comparable with statistical errors.
These are also shown in the lower panel of Fig.~\ref{fig:sin2Del},
where the phase shift $\sin^2 \delta$,
which is proportional to the scattering cross section
of the two-pion system, is plotted.
In Table~\ref{table:TanDel}
we see that the sign of $\delta$
at $\sqrt{s}< m_\rho$ ($m_\rho = 0.858 \pm 0.012$)
is positive (attractive interaction)
and that at $\sqrt{s}> m_\rho$
is negative (repulsive interaction) as expected.
This confirms the existence of a resonance
at a mass around $m_\rho$.

It may in principle be a straightforward task to estimate
the $\rho$ meson decay width
by fitting the phase shift data
with the Breit-Wigner formula.
The quark mass we worked with, however,
is much heavier than the realistic value,
so that a long extrapolation is needed.
Since we expect kinematic factors in the decay width
depend largely on the quark mass
while we made a simulation for only one set of the quark mass,
we avoid this direct measurement of the decay width
and take a different approach.
We parametrize the resonant behavior
of the $P$-wave phase shift in terms of
the effective $\rho\to\pi\pi$ coupling constant $g_{\rho\pi\pi}$
as
\begin{equation}
  \tan\delta
   = \frac{ g_{\rho\pi\pi}^2 }{ 6\pi }
     \frac{ k^3 }{ \sqrt{s} ( M_R^2 - s ) }
\ ,
\label{eq:tanD_g}
\end{equation}
as in the continuum theory,
where $M_R$ is the resonance mass and $g_{\rho\pi\pi}$ is defined by
\begin{equation}
  L_{\rm eff} = g_{\rho\pi\pi}
    \sum_{abc} \epsilon_{abc}
    ( k_1 - k_2 )_\mu  \ \rho_\mu^a(p) \pi^b(k_1) \pi^c(k_2)
\ .
\end{equation}
We may expect that such a coupling constant
does not vary too rapidly as the quark mass changes.

In (\ref{eq:tanD_g})
the invariant mass $\sqrt{s}$ and the momentum $k$
that appear in the effective theory in the continuum
satisfy the energy-momentum relations
in the continuum (\ref{eq:Disp_Two_Cont}).
A further discretization error may then appear
in the definition of $\sqrt{s}$ and $k$
upon the application of (\ref{eq:tanD_g})
to the phase shift calculated on the lattice.
We find, however, that
this does not lead numerically to a serious error.
In Table~\ref{table:TanDel}
we give the momentum $k_0^2 = s/4 - m_\pi^2$
calculated from $\sqrt{s}$,
and show that the difference between $k$ and $k_0$ is small.
Thus the error caused by the different definitions of the momentum
is negligible.
We adopt $k_0$ in the application of (\ref{eq:tanD_g}).

The results of the coupling $g_{\rho\pi\pi}$
and the resonance mass $M_R$ given by (\ref{eq:tanD_g}) are
\begin{eqnarray}
  g_{\rho\pi\pi}  &=&  6.25  \pm 0.67     \cr
  M_R             &=&  0.851 \pm 0.024    \cr
  M_R / m_\rho    &=&  0.992 \pm 0.033
\label{eq:FinalR_Cont}
\end{eqnarray}
using the energy-momentum relations
in the continuum (\ref{eq:Disp_Two_Cont}),
and
\begin{eqnarray}
  g_{\rho\pi\pi}  &=&  5.82  \pm 0.55     \cr
  M_R             &=&  0.906 \pm 0.028    \cr
  M_R / m_\rho    &=&  1.056 \pm 0.038
\label{eq:FinalR_Lat}
\end{eqnarray}
using those on the lattice (\ref{eq:Disp_Two_Lat}),
where $m_\rho$ is the $\rho$ meson mass
extracted from the time correlation function in the previous section.
In the lower panel of Fig.~\ref{fig:sin2Del}
we draw the curve for $\sin^2\delta$
given by (\ref{eq:tanD_g})
with $g_{\rho\pi\pi}$ and $M_R$
given in (\ref{eq:FinalR_Cont}) and (\ref{eq:FinalR_Lat}).
The position at $\sin^2\delta=1$
that corresponds to the resonance mass $M_R$
is also plotted in the upper panel of Fig.~\ref{fig:sin2Del}
for the two cases and compared with $m_\rho$.
We find that $M_R$ is consistent with $m_\rho$.

Assuming that the dependence of $g_{\rho\pi\pi}$
on quark mass is small,
we estimate the $\rho$ meson decay width
at the physical quark mass as
\begin{equation}
  \Gamma^{\rm ph}
    = \frac{ g_{\rho\pi\pi}^2 }{ 6\pi }
      \frac{ ( k^{\rm ph} )^3 }{ ( m_\rho^{\rm ph} )^2 }
    = g_{\rho\pi\pi}^2 \times 4.128 \ \ {\rm MeV}
\ ,
\end{equation}
where
$m_\rho^{\rm ph}=770\ {\rm MeV}$
is the actual $\rho$ meson mass
and $(k^{\rm ph})^2 = ( m_\rho^{\rm ph} )^2 /4 - (m_\pi^{\rm ph})^2$
($m_\pi^{\rm ph}=140\ {\rm MeV}$).
This yields
\begin{equation}
  \Gamma^{\rm ph} = 162  \pm 35  \ \ {\rm MeV}
\label{eq:FinalR_Gamm_Cont}
\end{equation}
for (\ref{eq:FinalR_Cont}),
and
\begin{equation}
  \Gamma^{\rm ph} = 140  \pm 27  \ \ {\rm MeV}
\label{eq:FinalR_Gamm_Lat}
\end{equation}
for (\ref{eq:FinalR_Lat}).
These estimates are consistent with experiment, $\Gamma = 150\ {\rm MeV}$.
The difference that arises from the two energy-momentum relations used
is comparable with the statistical error.
This is an encouraging result,
although we assumed that the coupling constant
does not depend on the quark mass to make a long extrapolation
and we did not trace the propagation of
discretization errors that appear in various steps so accurately.
%
%
\section{ Conclusions }
\label{Sec:Conclusions}
We have shown that a calculation of the $P$-wave scattering phase shift
for the $I=1$ two-pion system
and estimation of the decay width therefrom
are feasible on the lattice with present computing resources.
The phase shift data shows
the existence of a resonance at a mass
close to the vector meson mass obtained in the spectroscopy.
This resonance can certainly be identified with the $\rho$ meson.
We extracted the $\rho$ meson decay width from the phase shift data
and showed that it is consistent with the experiment.

We note, however, at the same time
several important issues that should be cleared in the future work.
One of the important issues is to reduce the discretization error
in the kinematic relations among the invariant mass $\sqrt{s}$,
the energy in the moving frame $W$ and the momentum $k$ on the lattice,
needed to evaluate the phase shift.
An obvious way to solve this problem
is to use a lattice closer to the continuum limit.

We have used the effective $\rho\to\pi\pi$ coupling constant
$g_{\rho\pi\pi}$ to extrapolate
from the point $m_\pi/m_\rho = 0.41$, where our simulation is made,
to the physical point $m_\pi/m_\rho = 0.18$,
assuming that $g_{\rho\pi\pi}$ does not depend on the quark mass.
More direct evaluation of the decay width is clearly desirable.
The decay width may be estimated directly
from the energy dependence of the phase shift data
by fitting the Breit-Wigner resonance formula,
if the simulations are made close to the physical quark mass
and we have data for several energy near the resonance mass.
We must leave these issues to studies in the future.
%
%
\section*{Acknowledgments}
This work is supported in part by Grants-in-Aid of the Ministry of Education
( Nos.
13135204, 
13135216, 
15540251, 
16540228, 
16740147, 
17340066, 
17540259, 
18104005, 
18540250, 
18740139  
).
The numerical calculations have been carried out
on VPP5000/80
at Academic Computing and Communications Center of University of Tsukuba.
%
%

%
%
\newpage
\appendix
%
%
%
%
%
%
%
%
%
%
\begin{table}[t]
\begin{ruledtabular}
\begin{tabular}{ l r r r r }
       & \multicolumn{1}{c}{ $n=1$ } &
       & \multicolumn{1}{c}{ $n=2$ } &  \\
\hline
$W^0_n$       &  $ 0.9805(51)$            &
              & \multicolumn{1}{c}{-----} &  \\
$\Delta W_n$  &  $-0.0441(60)$  &
              &  $ 0.105(22) $      \\
$W_n$         &  $ 0.9364(63)$  &
              &  $ 1.085(22) $      \\
\hline
        & \multicolumn{1}{c}{\bf Cont }
        & \multicolumn{1}{c}{\bf Lat  }
        & \multicolumn{1}{c}{\bf Cont }
        & \multicolumn{1}{c}{\bf Lat  } \\
$\sqrt{s}$       &  $ 0.7764(76)$    &   $ 0.8000(77) $
                 &  $ 0.951(25) $    &   $ 0.976(26)  $    \\
$k^2$            &  $ 0.0235(26)$    &   $ 0.0337(28) $
                 &  $ 0.099(12) $    &   $ 0.115(13)  $    \\
$k_0^2$  & \multicolumn{1}{c}{-----} &   $ 0.0328(27) $
         & \multicolumn{1}{c}{-----} &   $ 0.111(12)  $    \\
$\tan\delta$     &  $ 0.07906(93)$   &   $ 0.0736(66) $
                 &  $-0.38(11)   $   &   $-0.52(14)   $   \\
$\sin^2\delta$   &  $ 0.00621(14)$   &   $ 0.00539(96)$
                 &  $ 0.125(65)  $   &   $ 0.211(90)  $   \\
\end{tabular}
\end{ruledtabular}
\caption{
Energy of the two-pion system $W_n$
and the scattering phase shift $\delta$.
$W^0_1$ is the energy of the two free pions.
$W_n$ is reconstructed from the energy difference $\Delta W_n$
by $W_n = \Delta W_n + W^0_1$.
The invariant mass $\sqrt{s}$, the the momentum $k$
and the phase shifts $\delta$
calculated with the energy-momentum relations
in the continuum (\ref{eq:Disp_Two_Cont})
are referred to as {\bf Cont},
and those obtained with the relations
on the lattice (\ref{eq:Disp_Two_Lat})
are referred to as {\bf Lat}.
The momentum $k_0$ is defined by $k_0^2 = s/4 - m_\pi^2$.
All values with the mass dimension
are in units of the lattice spacing.
}
\label{table:TanDel}
\vspace{2cm}
\end{table}
%
\begin{table}[t]
\begin{ruledtabular}
\begin{tabular}{ l r r r r }
   & \multicolumn{1}{c}{ $\pi$  } &
   & \multicolumn{1}{c}{ $\rho$ } & \\
\hline
$m$  &  $ 0.3567(24)$  &  &  $ 0.858(12)$  \\
$E$  &  $ 0.6238(38)$  &  &  $ 1.019(27)$  \\
\hline
     & \multicolumn{1}{c}{\bf Cont }
     & \multicolumn{1}{c}{\bf Lat  }
     & \multicolumn{1}{c}{\bf Cont }
     & \multicolumn{1}{c}{\bf Lat  } \\
$p_{\rm eff}^2$      &  $ 0.2620(45)$   &   $ 0.2799(50)$
                     &  $ 0.301(59) $   &   $ 0.358(74) $   \\
$p_{\rm eff}^2/p^2$  &  $ 0.956(16) $   &   $ 1.021(18) $
                     &  $ 1.10(22)  $   &   $ 1.31(27)  $   \\
\end{tabular}
\end{ruledtabular}
\caption{
Mass $m$ and energy $E$
of the pion and the $\rho$ meson
with the momentum ${\bf p} = p {\bf e}_3 = (2\pi/L) {\bf e}_3$,
extracted from the time correlation function.
The momenta $p_{\rm eff}$
calculated from the dispersion relation
in the continuum (\ref{eq:Disp_One_Cont})
are referred to as {\bf Cont} and
that from the relation on the lattice (\ref{eq:Disp_One_Lat})
as {\bf Lat}.
All values are in units of the lattice spacing.
}
\label{table:Disp}
\end{table}
%
%
\end{document}